\documentclass[%
 reprint,
 amsmath,
 amssymb,
 aps,
 twocolumn,
 superscriptaddress
]{revtex4-1}

\usepackage{upgreek}

\usepackage{amsmath}
\usepackage{graphicx}
\usepackage{bm}
\usepackage[caption=false]{subfig}
\usepackage{float}
\usepackage{placeins}
\usepackage{xcolor}
\usepackage{siunitx}
\usepackage{titlesec}
\usepackage{gensymb}
\usepackage{xcolor}

\titleformat{\section}
    {\normalfont\itshape\filcenter}{\thesection}{1em}{}
\titlespacing*{\section}{0pt}{0.5\baselineskip}{\baselineskip}

\begin{document}

\preprint{APS/123-QED}

\title{Disorder-driven localization and electron interactions in Bi$_x$TeI thin films}

\author{Paul Corbae}
\affiliation{Department of Materials, University of California, Berkeley, California, 94720, USA}
\affiliation{Materials Science Division, Lawrence Berkeley National Laboratory, Berkeley, California, 94720, USA}
\thanks{To whom correspondence should be addressed; Email:pcorbae@berkeley.edu; fhellman@berkeley.edu}
\author{Nicolai Taufertshöfer}
\affiliation{Department of Physics, University of California, Berkeley, California, 94720, USA}
\affiliation{Physikalisches Institut, Julius-Maximilians-Universität Würzburg, Würzburg, D-97074, Germany}
\author{Ellis Kennedy}
\affiliation{Department of Materials, University of California, Berkeley, California, 94720, USA}
\affiliation{Materials Science Division, Lawrence Berkeley National Laboratory, Berkeley, California, 94720, USA}
\author{Mary Scott}
\affiliation{Department of Materials, University of California, Berkeley, California, 94720, USA}
\affiliation{Materials Science Division, Lawrence Berkeley National Laboratory, Berkeley, California, 94720, USA}
\author{Frances Hellman}
\affiliation{Department of Physics, University of California, Berkeley, California, 94720, USA}
\affiliation{Materials Science Division, Lawrence Berkeley National Laboratory, Berkeley, California, 94720, USA}

\date{\today}

\begin{abstract}
Strong disorder has a crucial effect on the electronic structure in quantum materials by increasing localization, interactions, and modifying the density of states. Bi$_x$TeI films grown at room temperature and \SI{230}{K} exhibit dramatic magnetotransport effects due to disorder, localization and electron correlation effects, including a MIT at a composition that depends on growth temperature.
The increased disorder caused by growth at 230K causes the conductivity to decrease by several orders of magnitude, for several compositions of Bi$_x$TeI. The transition from metal to insulator with decreasing composition $x$ is accompanied by a decrease in the dephasing length which leads to the disappearance of the weak-antilocalization effect. Electron-electron interactions cause low temperature conductivity corrections on the metallic side and Efros-Shklovskii (ES) variable range hopping on the insulating side, effects which are absent in single crystalline Bi$_x$TeI. The observation of a tunable metal-insulator transition and the associated strong localization and quantum effects in Bi$_x$TeI shows the possibility of tuning spin transport in quantum materials via disorder. 

\end{abstract}

\maketitle

\section*{Introduction}
Quantum materials such as topological materials or correlated electron phases have revealed remarkable emergent properties such as robust spin-momentum locked surface states and unconventional superconductivity, and have the potential to revolutionize technology with applications ranging from low power electronics to quantum computing. Many studies involve single crystals since disorder is typically seen as a drawback that hinders the emergence of interesting quantum properties. However, disorder can be useful in quantum materials and the observation of certain phenomena rely on the presence of disorder, such as the quantum anomalous Hall effect\cite{PhysRevMaterials.1.011201}. Quantum properties that exist in the amorphous structure  include superconductivity, magnetism, and topological phases, the latter seen via spin-momentum locked surface states in an amorphous analog of a 3D topological insulator \cite{Corbae2023}.

It is well understood that structural disorder can lead to electron localization, specifically via Anderson localization \cite{PhysRev.109.1492}. 
In these systems there is a finite density of states at the Fermi level, but the states are localized within a localization length that is typically much larger than the interatomic spacing. The Fermi energy is commonly tuned by composition and the correlation length diverges when the Fermi energy lies at the mobility edge \cite{nla.cat-vn1144667}. When the Fermi energy passes through the mobility edge, those states become delocalized. This transition is relatively well understood if interactions are not present which enables a single particle description. However, in systems where carriers are charged (electrons or holes), then Coulomb interactions are important and these introduce many body correlation effects. Additionally, disorder also increases these interactions, and in some cases interactions can dominate \cite{PhysRevLett.85.848}. The presence of interactions opens a Coulomb gap in the density of states at the Fermi level \cite{efrosshklov}. The interplay of strong disorder and interactions is an open field of research \cite{Moon2018}. 

Spin-orbit coupling (SOC) has a particularly strong effect on materials which are disordered, where localization can increase the strength of SOC relative to electron kinetic energy \cite{PhysRevB.90.125309}. On the metallic side of the metal-insulator transition (MIT), the presence of disorder leads to the suppression of the conductance due to the constructive quantum interference of time reversed paths in zero magnetic field and is known as weak-localization (WL). In the presence of strong SOC, the spins of time reversed paths rotate and destructively interfere leading to an enhancement of the conductance known as weak-antilocalization (WAL) \cite{RevModPhys.57.287,10.1143/PTP.63.707}. WAL and WL will manifest in the scattering lifetimes which can be studied via magnetotransport \cite{BERGMANN19841}. 
On the insulating side, where a fully formed Coulomb gap exists in the density of states, carriers exhibit variable range hopping (VRH) between sites where the hopping can be spin-dependent \cite{Reindl2019,Korzhovska2020}. In this regime, the application of a magnetic field can contract the localized wavefunction tails and lead to an increase in resistance, the so called wavefunction shrinkage model (WFS) \cite{Schoepe1988}.

\begin{figure*}
\centering
  \includegraphics[width=1\textwidth]{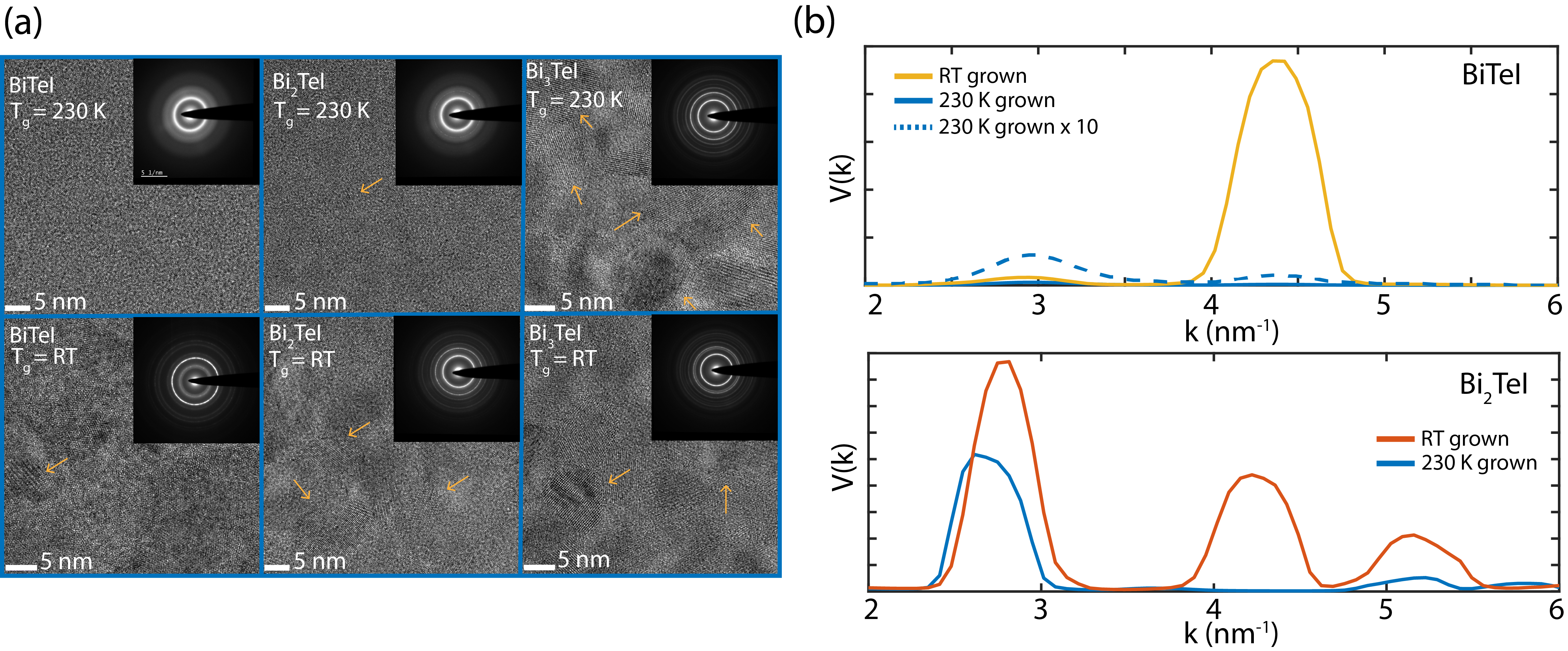}
  \caption{\textbf{Structural disorder in Bi$_x$TeI thin films.}  \textbf{(a)} HRTEM images Bi$_x$TeI for varying composition and growth temperature. Cold grown (\SI{230}{K}, LT) films are amorphous for low $x$ then nucleate crystallites with larger size as Bi concentration is increased. RT-BiTeI is amorphous with a few nanocrystals embedded in the matrix. RT growths $x=2-3$ are nanocrystalline showing crystallites larger in size compared to $x=1$. The volume fraction of nanocrystals is greater for $x=3$. Example nanocrystallites indicated by arrows. Inset of \textbf{(a)}, Diffraction patterns for Bi$_x$TeI for varying composition and growth temperature. LT-BiTeI is amorphous. LT-Bi$_2$TeI and LT-Bi$_3$TeI contain nanocrystallites in an amorphous matrix, with increasing nanocrystalline volume fraction for increasing $x$, evidenced also by the transition of broad to sharp rings with increasing $x$. RT-BiTeI has broad rings from small nanocrystals and the amorphous matrix, then as Bi concentration is increased there is an increase in the intensity of the rings. \textbf{(b)} Variance in the diffracted intensity as a function of scattering vector $k$ for RT and LT BiTeI, Bi$_2$TeI showing large differences in the relative peak heights at different scattering vectors $k$. The film grown at higher $T$ has more MRO and less disorder, evidenced by the large second peak in the variance. 
  }
  \label{fig:structure}
\end{figure*}

The crystalline Bi$_x$TeI system contains several distinct quantum materials with different emergent properties. BiTeI $(x=1)$ is a small gap Rashba semiconductor with a very large spin splitting \cite{Ishizaka2011}. BiTeI can be tuned by disorder or pressure to different ground states, specifically a topological insulator or Weyl semimetal phase \cite{PhysRevB.103.214203,Bahramy2012bitei}. Bi$_2$TeI $(x=2)$ is a weak topological insulator due to an even number of band inversions and a topological crystalline insulator from a mirror symmetry in the crystal structure leading to surface states on a distinct set of surfaces, both of which have been seen in experiment \cite{doi:10.1126/sciadv.aat0346,Avraham2020}. Finally, Bi$_3$TeI $(x=3
)$ is metallic, and has been proposed to be a topological metal \cite{Zeugner2017}. In BiTeI, the layers are van der Waals (vdW) bonded together, leading to a vdW gap. The Bi$_2$TeI and Bi$_3$TeI structures incorporate one and two Bismuth bilayers, respectively, in the van der Waals gap of BiTeI. Experimentally, single crystals of Bi$_x$TeI with $x=1,2,3$ show n-type metallic behaviour with $\rho(T)$ decreasing with decreasing temperature and $\rho \sim \SI{0.5}{m\Omega\cdot cm}$ at \SI{2}{K} \cite{https://doi.org/10.48550/arxiv.2209.02688,Ishizaka2011}. The metallic behaviour has been discussed as a result of non-stoichiometry leading to an n-type semiconductor for $x=1,2$. 
This Bi$_x$TeI system thus presents an opportunity to study the effects of disorder in a material with a wide range of topological and quantum states.

In this work we grow Bi$_x$TeI thin films for the first time at both room temperature (RT) and \SI{230}{K} (low temperature: LT) with varying levels of structural disorder to study how transport mechanisms are affected by disorder. Electron diffraction is used to characterize the structure and shows that both a reduced growth temperature and reduced Bi concentration $x$ cause increased structural disorder. Transport properties were measured as a function of temperature and magnetic field for an extensive range of compositions and disorder. A metal-insulator transition is seen with decreasing composition for both RT and LT samples. The increased structural disorder in the LT samples causes the MIT transition to occur at higher $x$ than in the RT samples. Temperature and magnetic field dependent transport provides evidence for weak antilocalization on the metallic side of the MIT, associated with strong spin-orbit coupling, which disappears with increased disorder and interaction effects in the LT samples. The dephasing length is extracted from the field dependence and decreases as a function of increased disorder with the dominant dephasing mechanism changing from electron-phonon interactions to electron-electron interactions. Our work demonstrates that strong disorder can localize states where SOC plays a strong role and increase interactions in the complex electronic structure in quantum materials, such as the Bi$_x$TeI family.

\section*{Results}

\begin{figure*}
\centering
  \includegraphics[width=1\textwidth]{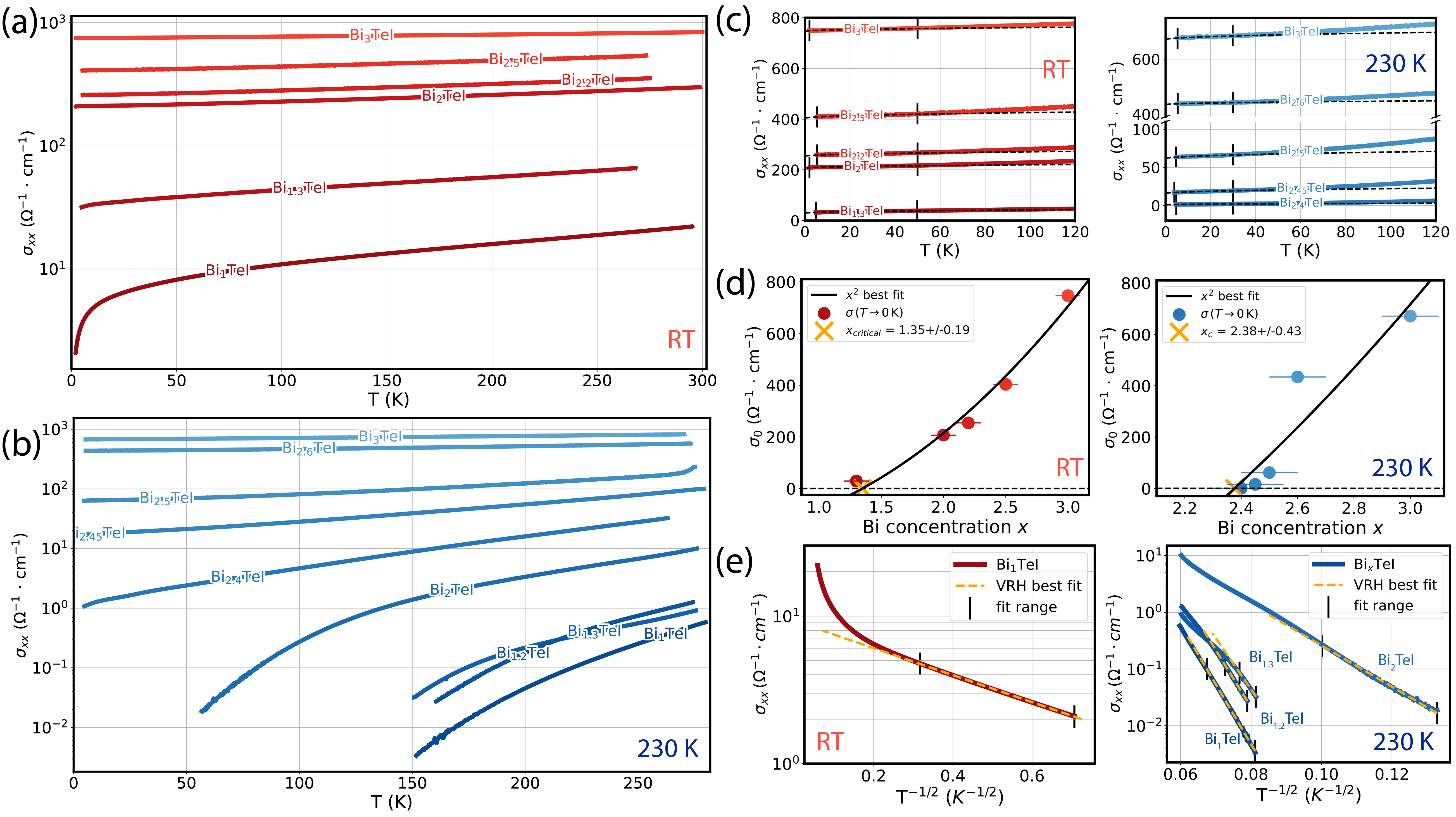}
  \caption{\textbf{Metal-insulator transition in Bi$_x$TeI for \bm{$x=1-3$}.} \textbf{(a)} Conductivity $\sigma$ vs. temperature for Bi$_x$TeI films grown at room temperature (RT) and \textbf{(b)} with substrates cooled down to $\approx$ \SI{230}{K} (LT). \textbf{(c-e)} left RT, right 230K (or LT). \textbf{(c)} Fits to the conductivity based on the presence of interactions, $\sigma(T) = \sigma_0 + aT^{1/2}$, where $\sigma_0 = \sigma(T=\SI{0}{K})$ and the $T^{1/2}$ term is due to Coulomb interactions which produce a precursor Coulomb gap in the density of states in a disordered system. \textbf{(d)} $\sigma_0$ vanishes as a power law for the metal-insulator transition approached from the metallic side for both RT and LT films. The critical Bi composition $x$ for RT grown is $x_c = 1.35 \pm 0.19$ and for LT films grown at \SI{230}{K} is $x_c = 2.38 \pm 0.43$. This significant difference in $x_c$ displays how the increased disorder in films grown at reduced temperature affects the electronic properties of the material, significantly increasing the localization of carriers. \textbf{(e)} VRH fits $\propto e^{-(T_0/T)^{\nu}}$ to Bi$_x$TeI films that show insulating behaviour. The fit range is indicated by the black bars. The hopping regime is well described with $\nu = 1/2$ indicating Coulomb interactions have to be taken into consideration and are of a relevant strength compared to the hopping energy.}
  \label{fig:MIT}
\end{figure*}

Growth temperature and composition both affect the structure. In Fig. \ref{fig:structure}(a), high-resolution transmission electron microscopy (HRTEM) real space images of low temperature-grown BiTeI (LT-BiTeI) shows the film is fully amorphous, while LT-Bi$_2$TeI has \SI{5}-\SI{10}{nm} crystalline precursors in an amorphous matrix evidenced by a few areas of limited lattice fringes with no clear boundary, and LT-Bi$_3$TeI is entirely nanocrystalline with nanocrystals $\sim \SI{10}-\SI{20}{nm}$. Room temperature BiTeI (RT-BiTeI) show a few nanocrystals of approximately \SI{5}{}-\SI{10}{nm} diameter embedded in an amorphous matrix.
In RT-Bi$_2$TeI and RT-Bi$_3$TeI, the diameter of the nanocrystals approximately double and the volume fraction increases with RT-Bi$_3$TeI being nearly entirely nanocrystalline. 
Inset in Fig. \ref{fig:structure}(a) shows parallel beam electron diffraction patterns for all of these samples, consistent with the characterization of disordered structure seen in HRTEM. Further structural data, including radially integrated intensity $I(k)$ of electron diffraction patterns, can be found in the supplemental information. The structure and degree of disorder for $x=1,2$ is thus greatly affected by growth temperature whereas for $x=3$ the structure is not significantly affected by growth temperature.

To further investigate the structure, fluctuation electron microscopy (FEM), a type of scanning nanodiffraction specifically suited for determining the degree of medium-range ordering (MRO) on the 1-4 nm length scale in amorphous materials, was used to measure the variance of the diffracted intensity $V(k)$ \cite{VOYLES2002147}. Regions of crystalline order produce large variations in the diffracted intensity from Bragg scattering, therefore the magnitude of $V(k)$ in amorphous samples is smaller than in the samples with a large nanocrystalline fraction. The relative peak heights of $V(k)$ between RT and \SI{230}{K} BiTeI, Bi$_2$TeI are shown in Fig. \ref{fig:structure}(b). The top panel of Fig. \ref{fig:structure}(b) compares $V(k)$ for the BiTeI films grown at RT and \SI{230}{K}, and shows that they have peaks at approximately the same $k$, consistent with them both being primarily amorphous as previously discussed, but the RT $V(k)$ has a significantly enhanced second peak, at \SI{4.4}{nm}$^{-1}$ due to the presence of nanocrystals. Similarly, RT-Bi$_2$TeI has a pronounced peak at \SI{4.4}{nm}$^{-1}$ compared to LT-Bi$_2$TeI, with a third peak resulting from a nanocrystalline population with ordering on the length scale of \SI{1.9}{\AA}. It is clear from Fig. \ref{fig:structure}(b) that a lower growth temperature reduces the magnitude of $V(k)$. In amorphous materials, $V(k)$ is used to understand the relative amount of MRO and at what length scales the ordering exists. The presence of peaks in the $V(k)$ of both LT-BiTeI and LT-Bi$_2$TeI show the amorphous films have MRO and the positions of the peaks indicate a degree of similarity between the bond lengths in both compositions. We will see that the samples that are more disordered with less MRO have greater resistivity.

\begin{figure*}
\centering
  \includegraphics[width=1\textwidth]{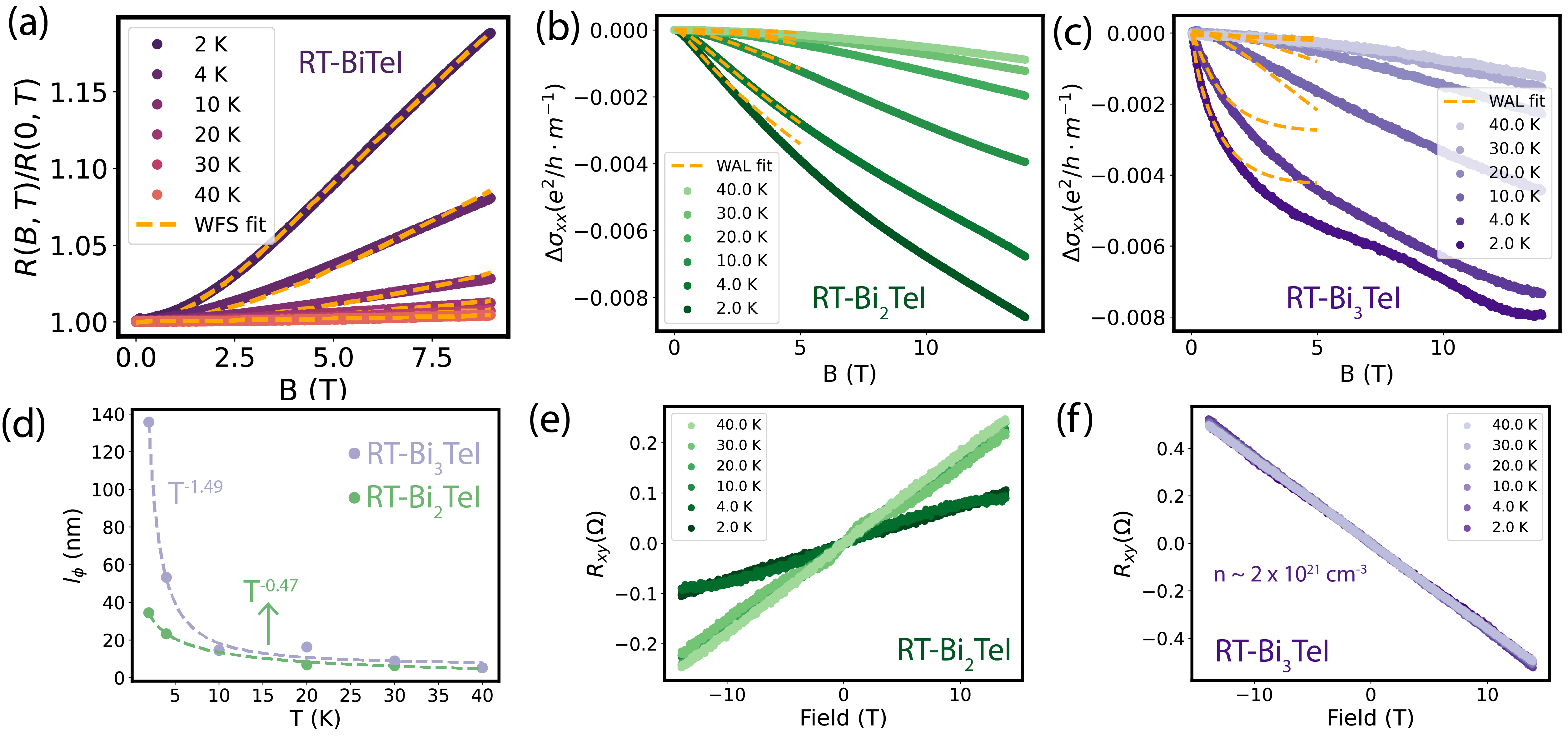}
  \caption{\textbf{Magnetotransport in RT Bi$_x$TeI thin films} \textbf{(a)} Normalized resistance of RT-BiTeI as a function of temperature and field. The film is highly resistive in the hopping regime and shows a positive MR that decreases with increasing measurement temperature. The low temperature date is fit well by the wavefunction shrinkage model. \textbf{(b)} Magnetoconductance of RT-Bi$_2$TeI, again negative and decreasing in magnitude with increasing measurement temperature but two orders of magnitude larger than in x=1.  Fits shown are 3D weak antilocalization.. \textbf{(c)} Magnetoconductance of RT-Bi$_3$TeI with 3D weak antilocalization fits. The sharp dip in the MC with field for \SI{2}{K} data is from the WAL effect. \textbf{(d)} Dephasing length versus temperature for RT-Bi$_2$TeI and RT-Bi$_2$TeI. The RT-Bi$_2$TeI data is fit best as $l_{\phi}$ proportional to $T^{-0.47\pm0.12}$ indicating dephasing due to electron interactions. For RT-Bi$_3$TeI the data is fit best as $l_{\phi}$ proportional to $T^{-1.49\pm0.16}$ indicating dephasing due to electron-phonon interactions. \textbf{(e)} Hall resistance for RT-Bi$_2$TeI. $R_{xy}$ is nonlinear and the Hall coefficient is positive. Below \SI{5}{K} the Hall coefficient is a factor of two smaller than above \SI{5}{K}. \textbf{(f)} Hall resistance for RT-Bi3TeI. $R_{xy}$ is linear, electron-like, and temperature independent. }
  
  \label{fig:MR}
\end{figure*}

Turning now to the electrical transport properties of these films, we find a transition from insulating to metallic behavior with increasing $x$, for both RT and LT samples. Fig. \ref{fig:MIT}(a,b) shows the conductivity of RT and LT Bi$_x$TeI samples versus temperature for a range of $x$, showing clear evidence of a MIT in Bi$_x$TeI for both RT and \SI{230}{K} grown films at different $x$. All samples are plotted here on log scales (a,b) or linear scales (c,d) of $\sigma$ vs $T$ in order to display the transition from metallic behavior at high $x$ to insulating behavior at lower $x$. The high $x$ samples (both RT and LT) show a nearly temperature independent conductivity, with $\sigma \sim 800$ $\Omega^{-1} \cdot$cm$^{-1}$, which increases slightly with increasing temperature, a common feature of amorphous metals above but near the MIT, and a sign of WL and quantum correction effects. We note again that for single crystal $x=3$, the $\sigma \sim 2000$ $\Omega^{-1} \cdot$cm$^{-1}$, and decreases slightly with increasing temperature while this disordered nanocrystalline sample shows a somewhat lower 
conductivity ($\sim 800$ $\Omega^{-1} \cdot$cm$^{-1}$) which slightly 
increases with increasing temperature, a signature of the localization effects of the disorder to be further analyzed by magnetotransport.  With decreasing $x$, $\sigma$ decreases and becomes increasingly temperature dependent, with increasingly sharp downturns seen at lower temperatures, particularly in the LT samples. The crystalline materials
are experimentally metals with conductivity that decreases with increasing temperature attributed to off-stoichimetry \cite{Ishizaka2011}.  By fitting the data to $\sigma_0 +AT^{1/2}$, an expression appropriate to metals near the MIT, values for $\sigma_0$ vs. $x$ are extracted and plotted in Fig. \ref{fig:MIT}(c), for RT and LT metallic samples where this fit is appropriate. These figures show $\sigma_0$ vanishing for $x$ near 1.5 (RT) and 2.4 (LT). For lower $x$ (lower $\sigma$) for both RT and LT films, the data is best viewed on a log scale, with an activated (exponential) temperature dependence, well fit by variable range hopping (VRH), as shown in \ref{fig:MIT}(e).  
LT samples with $x<2.4$ were not measured below \SI{50}{} or \SI{150}{K} (depending on $x$)  because their resistance exceeded M$\Omega$, the limit of the measurement setup.

We now look in greater detail at the temperature dependence of $\sigma(T)$. Altshuler and Aronov \cite{ALTSHULER19851} described the effect of Coulomb interactions between electrons in a disordered system which carve out a $\sqrt{E}$ gap in the density of states and cause a $T^{1/2}$ term in the conductivity for metallic systems, with a non-zero conductivity in the limit as T approaches zero. Fig. \ref{fig:MIT}(c) shows the low temperature fit including a $T^{1/2}$ term for both warm and cold grown samples on the metallic side. Curves with nonzero $\sigma_0 = \sigma(T= \SI{0}{K})$ allows us to make an estimate of the composition $x$ of Bi$_x$TeI films where the MIT occurs. From scaling theory of localization it is assumed that $\sigma_0$ vanishes continuously when being approached from the metallic side with a critical exponent $\beta$ such that $\sigma_0 \propto (x-x_c)^{\beta}$. In literature, a scatter of results for $\beta$ ranging from 0.5 to 2 is found, and a debate about the interpretation of the transition is ongoing \cite{AMacKinnon1994}. 
$\sigma_0$ values for both warm and cold grown films were fit to this expression, with a best fit value of $\beta = 2$ shown in Fig. \ref{fig:MIT}(d). From this data we report a critical composition $x_c$ of $x_{c,warm} = 1.35 \pm 0.19$ and $x_{c,cold} = 2.38 \pm 0.43$. 

On the insulating side of the MIT, localized states contribute to conductivity at non-zero temperature by hopping to states near the Fermi level. $\sigma$ is proportional to $e^{-(T_0/T)^{\nu}}$ where $T_0$ is a temperature which can be related to the localization length (length scale over which the amplitude of the electron wavefunction decays) and $\nu=1/4$. Efros and Shklovskii showed that when Coulomb interactions are taken into account the exponent is $\nu = 1/2$ \cite{efrosshklov}. $\nu = 1/4$ is only obeyed when the in-gap density of states is small but not zero, as in amorphous semiconductors, leading to the Mott band being larger than the Coulomb gap \cite{efrosshklov}. Fig. \ref{fig:MIT}(e) shows curves with insulating character ($x < x_c$) for warm and cold grown films where the fits indicate the hopping regime. In the warm grown films only RT-BiTeI is insulating and is well described by a VRH model with $\nu = 1/2$ up to \SI{10}{K}, indicating Coulomb effects are important. The cold grown samples with a composition below $x_{c,cold}$ show highly insulating behaviour which can be described as well within the Efros and Shklovskii picture of significant Coulomb interactions with an exponent of $\nu = 1/2$, not only for the compositions close to the MIT.

The magnetoresistance is a probe of the different dephasing mechanisms from the insulating (hopping) to metallic (diffusive) regimes in RT grown Bi$_x$TeI films. Low temperature magnetoresistance measurements are sensitive to quantum corrections in the diffusive regime such as WAL or WL which manifest as positive or negative MR (negative or positive magnetoconductance). 
Fig. \ref{fig:MR} shows magnetoresistance (MR) and magnetoconductance (MC) curves as a function of temperature and field. 
RT-BiTeI, which is an insulator, has a positive, linear MR even when it is in the hopping regime where previous works on both crystalline and amorphous spin-orbit materials have shown negative magnetoresistance in the hopping regime \cite{Reindl2019,Korzhovska2020,NMPounder_1991}. The MR for RT-BiTeI is shown in \ref{fig:MR}(a) with a WFS fit since it is in the hopping regime where the 3D WAL theory is not applicable. The WFS fit at \SI{2}{K} gives an excellent fit from \SI{0}{} to \SI{9}{T}. RT-Bi$_2$TeI shows the onset of WAL with a larger decrease of the MC with applied field compared to RT-BiTeI. RT-Bi$_3$TeI shows a sharp, clear WAL cusp indicated by the rapid decrease of the MC at small magnetic fields. 
More quantitative estimations of WAL effects can be made by fitting our data to the three dimensional theory of WAL \cite{Nakamura2020} (details presented in supplemental information).   
The fits to the 3D WAL theory for RT-Bi$_2$TeI and RT-Bi$_3$TeI are shown in Fig. \ref{fig:MR}(b,c). $l_{\phi}$ rises to values of \SI{35}{nm} and then \SI{136}{nm} in the progressively more ordered RT-Bi$_2$TeI and RT-Bi$_3$TeI respectively (both magnitudes larger than $l_{SO}$), explaining the emergence of WAL for increasing $x$.

The dependence of the dephasing length $l_{\phi}$ on temperature gives insights into the dominant dephasing mechanisms. 
The dephasing length in RT-Bi$_2$TeI follows a $T^{-0.47\pm0.12}$ scaling,  Fig. \ref{fig:MR}(d). If the dominant dephasing mechanism were electron-phonon interactions in this 3D disordered system, $p$ would be between 1 and 3 \cite{JJLin_2002}.  Since $p < 1$, for RT-Bi$_2$TeI, the dephasing mechanism is  nearly certainly due to electron-electron interactions.  By contrast, in 
RT-Bi$_3$TeI, $l_{\phi}$ follows the scaling of $T^{-1.49\pm0.16}$ indicating the dephasing mechanism is dominated by electron-phonon interactions \cite{RevModPhys.57.287}, Fig. \ref{fig:MR}(d).

Hall effect data on RT samples is shown in Fig. \ref{fig:MR}(e,f). Since RT-BiTeI is in the hopping regime, we don't extract a carrier concentration because the Hall effect is difficult to interpret in this regime, however the Hall voltage is positive, linear, and temperature independent. 
Below \SI{5}{K}, the RT-Bi$_2$TeI Hall resistance ($R_{xy}$) is positive in field but non-linear in $B$
signaling multiple types of carriers. The Hall coefficient in RT-Bi$_2$TeI is temperature dependent and below \SI{5}{K} is a factor if two smaller than above \SI{5}{K}; this can be a linked to the presence of electron interactions since $\delta R_H /R_H = 2\delta R /R$ \cite{PhysRevB.22.5142}. The ratio of $(\delta R_H /R_H)/(\delta R /R)$ gives 2.3 for RT-Bi$_2$TeI. For this reason we do not extract a carrier density as the Drude picture is based on a single particle picture. RT-Bi$_3$TeI shows a temperature independent, negative (electron-like) carrier concentration of $n_{3D} \sim 2 \times 10^{21}$ \SI{}{cm^{-3}}.

\begin{figure}
\centering
  \includegraphics[width=0.42\textwidth]{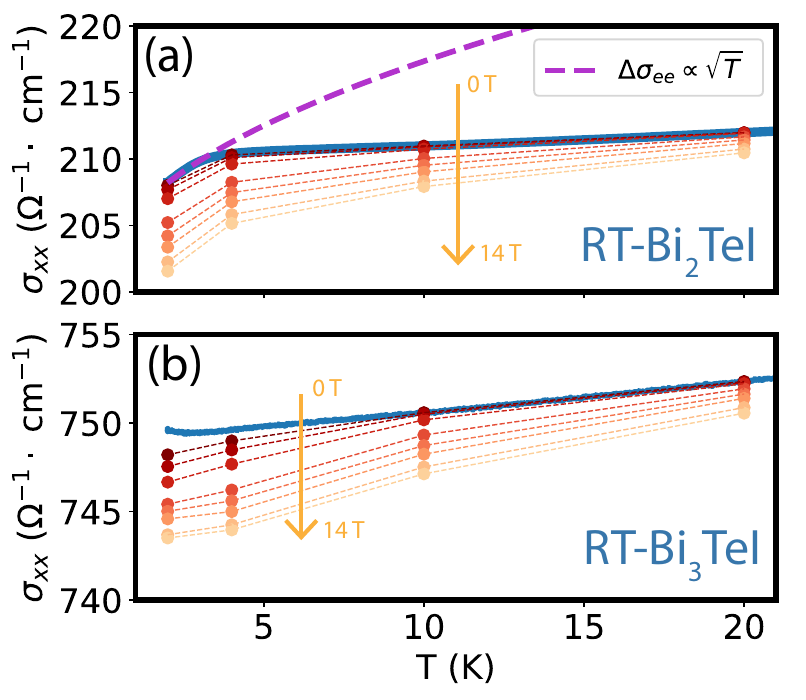}
  \caption{\textbf{Electron interactions in RT Bi$_x$TeI thin films} \textbf{(a)} $\sigma_{xx}$ vs $T$ for different applied fields in RT Bi$_2$TeI films. The low temperature conductivity decreases with increasing applied field as quantum interference is destroyed and the dip below \SI{5}{K} scales as $\sqrt{T}$, indicating electron interactions dominate below 5K, and do not change with magnetic field. Blue line is zero-field data which decreases with decreasing temperature, an indication that WAL is not relevant.. The purple dashed line shows the low temperature conductivity modifications from electron interactions. \textbf{(b)} $\sigma_{xx}$ vs $T$ for different applied fields in RT Bi$_3$TeI films. Blue line is zero-field data, which increases with decreasing temp (below~2K) due to WAL. Applying a magnetic field eliminates WAL, such that the conductivity decreases with decreasing temperature at all temperatures in fields the application of a field destroys this effect and the the conductivity decreases for all temperatures.}
  \label{fig:eei}
\end{figure}

The role of enhanced correlations from disorder was studied further in Bi$_x$TeI films on the metallic side of the MIT in both RT and LT films. The application of a magnetic field will suppress quantum interference effects but not electron interactions since interaction effects occur through a particle-hole diffusion channel that is insensitive to magnetic flux \cite{RevModPhys.57.287}. 
Fig. \ref{fig:eei}(a) shows the RT-Bi$_2$TeI samples have a low temperature (below 4K) field-independent decrease in the conductivity and a positive MR. These results show that electron-electron interactions dominate at the lower temperatures, with the field-dependent positive MR due to WAL and not WL, which is consistent with the value of the dephasing length power law exponent. 
The RT-Bi$_3$TeI sample (Fig. 4b) in zero field  shows a low temperature upturn in the conductivity which is a result of WAL which is suppressed by a magnetic field; this positive MR effect is visible up to 20K, as for RT-Bi2TeI shown in Fig. 4a.. There are no low temperature interaction corrections to the conductivity in contrast to Bi$_2$TeI.
The observation of electron interactions in RT-Bi$_2$TeI and not RT-Bi$_3$TeI indicates that electron interactions emerge as $x$ approaches $x_c$ for the MIT in RT-Bi$_x$TeI.

\section*{Discussion}

One of the more striking results is that the amorphous (grown at \SI{230}{K}) LT-BiTeI has a conductivity that is orders of magnitude smaller than RT-BiTeI, and RT-BiTeI has a conductivity that is orders of magnitude smaller than single crystal BiTeI, clear evidence for the very large impact that disorder has on the conductivity of this material. 
Similar orders of magnitude differences are seen between LT-Bi$_2$TeI and RT-Bi$_2$TeI and the single crystal Bi$_2$TeI, showing that increased structural disorder dramatically increases carrier localization. For Bi$_3$TeI, all three (LT, RT and single crystal) are metallic, with similar conductivity, which is perhaps unsurprising since the disorder, even for LT Bi$_3$TeI, is moderate.

The strong SOC material Bi$_x$TeI undergoes a disorder-driven metal-insulator transition with varying $x$ that is dependent on the structure. The metal-insulator transition is accompanied by weak antilocalization on the metallic side and hopping transport on the insulating side. Structural disorder leads to increased electron interactions as evidenced by Efros-Shklovskii hopping and a reduction in the low temperature conductivity on the metallic side. Power-law fits suggest the dephasing mechanisms result from electronic correlations, consistent with the observed conductivity as a function of temperature and field.

\section*{Conclusion}

The observed metal-insulator transition as a function of composition and of growth temperature shows that structural disorder can be strong enough to localize carriers in materials which have complex bandstructures such as Rashba or topological materials, unlike in simple metals where the disorder of even a fully amorphous material is insufficient to localize the carriers. This transition happens when the Bi$_x$TeI films become predominantly amorphous showing the localization happens in a regime where Bloch states begin to become unsuitable description of the electronic states, evidenced by the good fit of the wavefunction shrinkage model and variable range hopping, indicating the wavefunctions become localized with exponentially decaying tails. Our study shows that growth temperature can be used to tune the structural disorder in quantum materials. Further studies of the local environment would provide insight into the details of localization. We observed increased electron-electron interactions with structural disorder. By introducing correlations with the quantum phases of Bi$_x$TeI such as Rashba or topological states, these materials can be used to study the convergence of many-body physics and topological phases by tuning disorder from the single crystal. 
Our work highlights how structural disorder localizes carriers and tunes functionality in quantum materials with strong SOC.
We expect our work to motivate an effort to utilize disorder in quantum materials, enabling materials discovery that can provide a path towards scalable quantum devices.

\section*{Methods}
We grew \SI{100}{nm} thin films of Bi$_x$TeI with various Bi concentrations on a-SiN$_x$ covered Si substrates at room temperature (RT) and \SI{230}{K} (LT). The films were grown out of three effusion cells (Bi,Te,BiI$_3$) in a UHV chamber with a base pressure of $10^{-10}$ \SI{}{Torr}. Different compositions were achieved by varying the Bi rate relative to the Te and I rate. 
The film thicknesses were confirmed using profilometry and the compositions were determined by energy dispersive spectroscopy (EDS) and x-ray photoelectron spectroscopy (XPS). The structure of the room temperature and cold grown Bi$_x$TeI films was characterized using high resolution TEM, parallel beam diffraction, and scanning nanodiffraction. For transport measurements the films were patterned into Hall bar devices ($l=$ 600 $\mu$m, $w=$ 200 $\mu$m) using conventional photolithography techniques. The films were measured in a custom closed cycle cryostat and a Quantum Design PPMS using standard four-point DC and low frequency lock-in techniques.
We used pre-patterned thin Cr/Au (\SI{2}{nm}+\SI{8}{nm}) contacts on the substrates on which the films were grown and subsequently ex-situ patterned into Hall bars using lift-off techniques. The structure and electronic properties of this material is badly influenced by any sort of annealing, which induces crystallization. The choice of pre-made Cr/Au contact pads and a lift-off technique was made to eliminate any annealing. Large Cr/Au contact pads were used to reduce contact resistance which can be significant on highly insulating films.

\section*{Acknowledgements}
This work was
primarily funded by the US Department of Energy, Office of
Science, Office of Basic Energy Sciences, Materials Sciences
and Engineering Division under Contract No. DE-AC02-05-
CH11231 (NEMM program MSMAG). Work at the Molecular Foundry was supported by the Office of Science, Office of Basic Energy Sciences, of the U.S. Department of Energy under Contract No. DE-AC02-05CH11231. We thank Alex Liebman-Pelaez and J.G. Analytis for use of their PPMS. We thank Maximilian Hofer for help implementing a fitting procedure of the Zeta function. 

\section*{Author contributions}
This project was designed by P.C. and F.H. The samples were grown by P.C. and N.T. The TEM was done by E.K. P.C. and N.T. performed transport measurements. N.T. performed the lithography. P.C., N.T., E.K., and F.H. all took part in the data analysis. The manuscript was written by P.C., N.T., E.K., M.H., and F.H. All authors contributed in discussing the results and forming conclusions. P.C. and N.T. contributed to this work equally.

\bibliographystyle{apsrev4-2}
\bibliography{bib.bib}

\end{document}